\documentclass[twocolumn,showpacs,preprintnumbers,amsmath,amssymb,nofootinbib]{revtex4-1}
\usepackage{graphicx}% Include figure files
\usepackage{dcolumn}% Align table columns on decimal point
\usepackage{bm}% bold math
\usepackage{amsfonts}

\def\bequ{\begin{equation}}
\def\eequ{\end{equation}}
\def\be{\begin{equation}}
\def\ee{\end{equation}}

\begin{document}

%\preprint{APS/123-QED}

\title{Ergo-spheres, ergo-tori and ergo-Saturns \\ for Kerr black holes with scalar hair}

\author{Carlos Herdeiro}
%\email{herdeiro@ua.pt}
\author{Eugen Radu}
\email{herdeiro@ua.pt; eugen.radu@ua.pt}
\affiliation{\vspace{2mm}Departamento de F\'\i sica da Universidade de Aveiro and I3N \\
Campus de Santiago, 3810-183 Aveiro, Portugal \vspace{1mm}}%

\date{May 2014}

\begin{abstract}
We have recently reported the existence of Kerr black holes with scalar hair in General Relativity minimally coupled to a massive, complex scalar field~\cite{Herdeiro:2014goa}. These solutions interpolate between boson stars and Kerr black holes. The latter have a well known topologically $S^2$ ergo-surface (\textit{ergo-sphere}) whereas the former develop a $S^1\times S^1$ ergo-surface (\textit{ergo-torus}) in a region of parameter space. We show that hairy black holes always have an ergo-region, and that this region is delimited by either an ergo-sphere or an \textit{ergo-Saturn} -- i.e. a $S^2\oplus (S^1\times S^1)$ ergo-surface. In the phase space of solutions, the ergo-torus can either appear disconnected from the ergo-sphere or pinch off from it. 
We provide a heuristic argument, based on a measure of the size of the ergo-region, that superradiant instabilities - which are likely to be present - are weaker for hairy black holes than for Kerr black holes with the same global charges. We observe that Saturn-like, and even more remarkable, ergo-surfaces should also arise for other rotating `hairy' black holes.
\end{abstract}

\maketitle

%%%
%%%%%%%%%%%%%%%%%%%%%%%%%%%%%%%%%%%%%%%%%%%%%%%%%%%%%%%%%%%%%%%%%%%%%%%%%%%%%%
%\noindent{\bf{\em I Introduction.}}
%%%%%%%%%%%%%%%%%%%%%%%%%%%%%%%%%%%%%%%%%%%%%%%%%%%%%%%%%%%%%%%%%%%%%%%%%%%%%%

%%%%%%%%%%%%%%%%%%%%%%%%%%%%%%%%%%%%%%%%%%%%%%%%%%%%%%%%%%%%%%%%%%%%%%%%%%%%%%
\section{Introduction}
%%%%%%%%%%%%%%%%%%%%%%%%%%%%%%%%%%%%%%%%%%%%%%%%%%%%%%%%%%%%%%%%%%%%%%%%%%%%%%
 Kerr black holes (BHs) can support regular scalar hair~\cite{Herdeiro:2014goa}. The existence of these hairy black holes (HBHs), continuously connected to Kerr BHs, can be understood from the fact that real frequency bound states exist for a massive, complex, Klein-Gordon scalar field on the Kerr background. Such bound states may, moreover, be understood as zero modes, from the viewpoint of the superradiant instability that afflicts the Kerr solution in the presence of the aforementioned scalar field~\cite{Press:1972zz,Damour:1976kh,Zouros:1979iw,Detweiler:1980uk,Hod:2012px,Hod:2013zza,Cardoso:2013krh,Shlapentokh-Rothman:2013ysa}. 

The existence of such solutions seems surprising at first sight, in view of a number of no-(scalar-)hair theorems for four dimensional, regular, asymptotically flat, BHs~\cite{Bekenstein:1996pn}. HBHs circumvent such theorems due to a phase-like time dependence for the scalar field. Actually, a no-(scalar-)hair theorem was proposed in~\cite{Pena:1997cy} with precisely this time dependence for \textit{static} BHs. But including rotation is mandatory for the existence of these HBHs~\cite{Herdeiro:2014ima}, making them consistent with the result in~\cite{Pena:1997cy}. The phase-like time dependence does not survive, however, at the level of the line element. As such the geometry is stationary and axi-symmetric, but the full solution is invariant under the action of a single helicoidal Killing vector field. The existence of BHs with this property was first appreciated in five dimensional Anti-de Sitter space~\cite{Dias:2011at}.  

HBHs are more star-like than Kerr BHs. This means their physical properties can be markedly different from those of the Kerr solution, as illustrated in~\cite{Herdeiro:2014goa} for both the quadrupole moment and the orbital frequency at the innermost stable circular orbit. As such HBHs can be used to bench mark deviations from the Kerr geometry, and to construct templates, for observable quantities, exhibiting such deviations. The templates can then be compared with future observations of BHs, for instance, with the Event Horizon Telescope~\cite{Loeb:2013lfa}. 

It is conceivable, moreover, that HBHs may play a role in astrophysics. To probe this possibility we are led to considering their stability. In~\cite{Herdeiro:2014goa} it was shown HBHs are entropically favored to Kerr BHs, when both solutions co-exist for the same global charges -- i.e. in the region of non-uniqueness. Despite this indication of stability, HBHs may be afflicted by instabilities, namely the superradiance associated to the massive scalar field they support. Indeed, HBHs connect to Kerr BHs precisely at the threshold of the superradiant instability of a particular mode, with a given azimuthal harmonic index $m$. Thus superradiant instabilities caused by modes with higher $m$ may be present. 

Studies of the scalar superradiant instabilities for the Kerr case may be performed simply by computing the Klein-Gordon equation in the Kerr background. By contrast, since HBHs have a non-vanishing scalar field, a study of scalar field perturbations necessarily involves the coupled scalar-gravitational perturbations (see \cite{Cardoso:2007az} for such a study applied to boson stars, within some approximations). Before attempting such challenging computation -- in particular in view of the numerical nature of the solutions -- it may be informative to understand the structure of the ergo-region of HBHs, since the latter is intimately connected to the superradiant instability. Such study will be presented in this paper. 

HBHs interpolate between Kerr BHs and boson stars. The former have a well known (topologically) $S^2$ ergo-surface -- the \textit{ergo-sphere} -- whereas the latter develop a $S^1\times S^1$ ergo-surface -- an \textit{ergo-torus} -- in some region of phase space~\cite{Kleihaus:2007vk}. We shall show that the ergo-surface for HBHs is a composition of these two building blocks.  In a large region of the parameter space there is only an ergo-sphere; in the remaining part, an ergo-torus is also present. The latter may emerge disconnected from the ergo-sphere or may pinch-off from it. In either case the topology of the total ergo-surface of the solution changes from $S^2$ to $S^2\oplus (S^1\times S^1)$; we dub the latter ergo-surface as an \textit{ergo-Saturn}.  

To the best of our knowledge, HBHs provide the first example of such ergo-surfaces in four dimensional asymptotically flat geometries. We remark, however, that
toroidal and Saturn-like ergo-surfaces are known in other contexts, including  black rings~\cite{Emparan:2001wn} and black Saturns~\cite{Elvang:2007rd} in higher dimensional gravity (see~\cite{Elvang:2008qi} for a discussion of ergo-region mergers in higher dimensions and~\cite{Costa:2009wj} in interacting Kerr BHs). Also, toroidal ergo-regions have been found in four dimensional  -- but non-asymptotically flat -- magnetized BHs~\cite{Gibbons:2013yq}. 

The existence of ergo-regions indicates that HBHs will be afflicted by superradiant instabilities. A proof of existence of such instabilities and a computation of their time scales requires a detailed analysis, as discussed above. We shall not pursue such computation herein; we shall, however, give an heuristic argument  that in the region of phase space where an unambiguous comparison can be made with Kerr BHs -- the region where there is non-uniqueness -- superradiant instabilities are \textit{weaker} for HBHs. The argument relies on making quantitative the intuitive notion that the `size' of the ergo-region should be positively correlated with the strength of the instability. We suggest that a quantitative measure of such `size' is provided by a normalized difference between the area of the total ergo-surface $A_E$ and the area of a spatial section of the event horizon $A_H$:
\bequ
\Delta a\equiv \frac{A_E-A_H}{16\pi M^2} \ ,
\label{ergosize}
\eequ
where $M$ is the ADM mass. This quantity -- which we dub \textit{ergo-size} -- is well defined even in the absence of an event horizon, as for the case of rotating boson stars. After arguing that $\Delta a$ is an appropriate measure of the strength of the instability for the Kerr-Newman family --  in the regime where the Compton wave length of the field is much larger than the BH --  we will show that $\Delta a$ is smaller for HBHs than for Kerr solutions with the same mass and angular momentum. Such observation suggests superradiant instabilities will be weaker for the former than for the latter.

This paper is organized as follows. In Section \ref{seceq} we shall make a scan of ergo-surfaces in the phase space of solutions, describing precisely where ergo-spheres and ergo-Saturns arise and how the transition between them occurs. In Section \ref{stability} we first review the time scales for scalar superradiant instabilities in Kerr and rephrase them in terms of the quantity $\Delta a$ defined above. Then we compare $\Delta a$ for Kerr and HBHs and conclude some lower limits for the e-folding times of superradiant instabilities for HBHs. In Section \ref{discussion}  we briefly discuss other models where ergo-Saturns --  and even more involved ergo-surfaces -- should occur. 

%%%%%%%%%%%%%%%%%%%%%%%%%%%%%%%%%%%%%
\section{Ergo-surfaces}
\label{seceq}
%%%%%%%%%%%%%%%%%%%%%%%%%%%%%%%%%%%%%
The scalar HBHs found in~\cite{Herdeiro:2014goa} are asymptotically flat, regular (on and outside an event horizon) solutions of the Einstein-Klein-Gordon system,  described by:
\begin{eqnarray}
\label{ansatz}
ds^2 &=&e^{2F_1}\left(\frac{dr^2}{N }+r^2 d\theta^2\right)+e^{2F_2}r^2 \sin^2\theta (d\varphi-W dt)^2
\nonumber \\
&&-e^{2F_0} N dt^2, \qquad
~~{\rm with}~~N=1-\frac{r_H}{r},
\end{eqnarray}
\begin{eqnarray}
\Psi=\phi(r,\theta)e^{i(m\varphi-w t)},
\label{ansatz2}
\end{eqnarray}
where $F_i,W,\phi$, $i=0,1,2$, are real functions of $(r,\theta)$ that are found numerically; explicit examples will be provided in~\cite{Herdeiro:2014}. Also, $w>0$ is the scalar field frequency and $m=\pm 1,\pm 2$\dots is the azimuthal harmonic index. Finally, $r_H$ corresponds to the radial coordinate of the BH horizon. Putting $r_H=0$ one obtains solutions without a horizon, describing rotating boson stars. HBHs can have different numbers of nodes of the scalar profile $\phi(r,\theta)$. As in~\cite{Herdeiro:2014goa},   we focus on nodeless solutions with $m=1$ and even parity, which are classified by the ADM mass $M$, angular momentum $J$ and normalized Noether (scalar) charge $q$.

\begin{figure}[h!]
\centering
\includegraphics[height=2in]{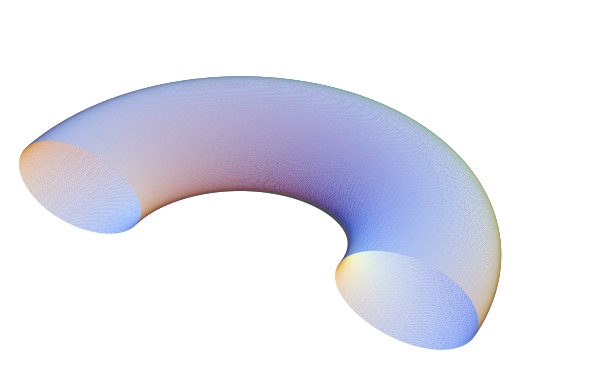}\\
\includegraphics[height=2.3in]{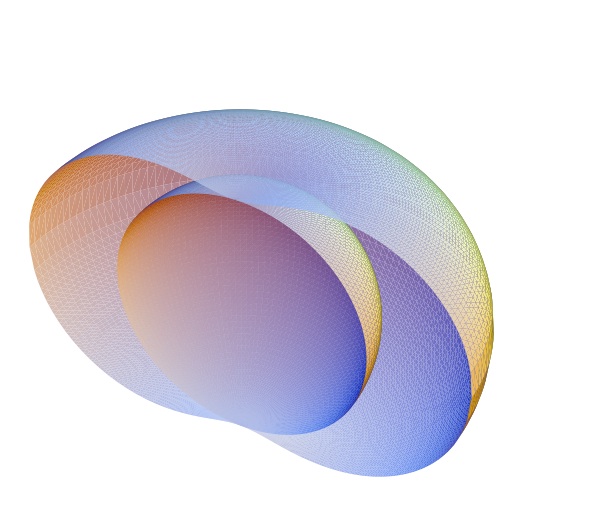}\\
\includegraphics[height=2.4in]{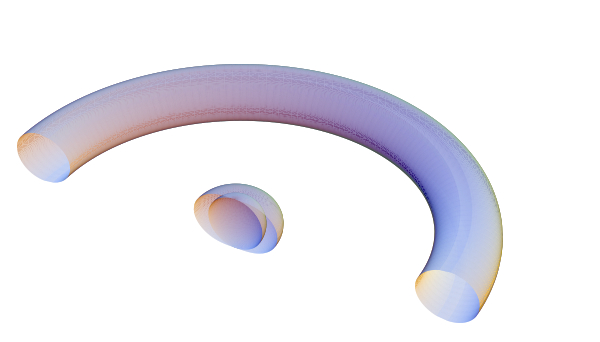}\\
\caption{Ergo-surfaces for a type II boson star (top) with physical parameters $(M,J,q)=(0.811,0.626,1)$, a type I HBH (middle) with parameters $(0.708,0.494,0.804)$  and a type III HBH (bottom), with $(1.007,0.876,0.986)$. In the bottom two plots, the event horizon is also exhibited (as the innermost spherical surface). The figures are not embedding diagrams; they use the numerical data obtained in the coordinates of eq. (\ref{ansatz}), but plotted as standard spherical coordinates.  Also the overall scale is different for each panel. As such, these figures should be regarded simply as illustrations.} 
\label{3dergo}
\end{figure}

Ergo-surfaces are invariantly defined as the locus at which the normalized time-like Killing vector field at infinity $\xi=\partial_t$ becomes null:
\begin{equation}
\xi\cdot \xi=0  \ \ \ \Leftrightarrow\  \ \ 0=g_{tt}=-e^{2F_0} N+W^2e^{2F_2}r^2 \sin^2\theta  \ .
\end{equation}
Let us consider the last equation for two cases: 
\begin{itemize} 
\item[i)] for HBHs, $r_H\neq 0$; then, at the horizon ($r=r_H$), $g_{tt}$ is positive on the equatorial plane ($\theta=\pi/2$). Since asymptotic flatness requires $g_{tt}\left|_{r=\infty}=-1\right.$, then $g_{tt}$ must have an \textit{odd} number of zeros between the horizon and infinity along the equatorial plane. Along the symmetry axis, on the other hand, $g_{tt}$ vanishes on the horizon and therefore it must have an even number of zeros outside the horizon.
\item[ii)] for boson stars, $r_H=0$; along the symmetry axis, $\theta=0, \pi$ -- in particular for $r=0$ -- $g_{tt}$ is negative. Asymptotic flatness then requires that $g_{tt}$ must have an \textit{even} number of zeros (along any $\theta={\rm constant}$ plane).
\end{itemize}
The upshot of this simple reasoning is that HBHs \textit{must have an ergo-region} and boson stars cannot have an ergo-sphere. Precisely what happens requires a more detailed analysis. Performing such an analysis, we find that boson stars can either have no ergo-region at all - dubbed \textit{Type 0 boson stars} - or have a toroidal ergo-region -- \textit{Type II boson stars}. On the other hand, HBHs can have only an ergo-sphere -- which we call \textit{Type I HBHs} -- or an ergo-Saturn --  \textit{Type III HBHs}. The type number, therefore, corresponds to the number of zeros of $g_{tt}$ along the equatorial plane, from either the horizon or the origin (in the case of boson stars). These different types of ergo-surfaces are illustrated in Fig. \ref{3dergo} as 3D plots obtained from our numerical data. 
%We emphasize the plots are not embedding diagrams; they use the numerical data obtained in the coordinates of eq. (\ref{ansatz}), but plotted as standard spherical coordinates. 

\begin{figure}[h!]
\centering
\includegraphics[height=2.48in]{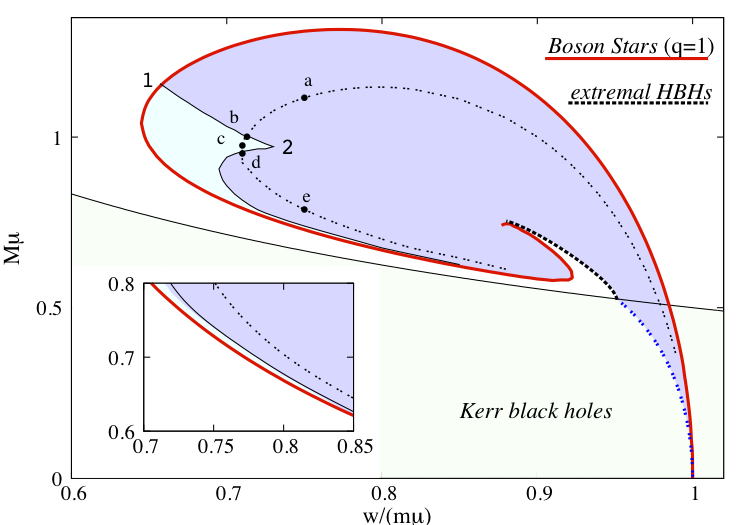}\\
\caption{Domain of existence of HBHs for $m=1$ in $M$-$w$ space (shaded region). The dotted black curve corresponds to solutions with $q=0.985$. Along this curve we have selected five points, labelled as {\bf a} to {\bf e}. 
%The ergo-surfaces for these five solutions are displayed in Fig. \ref{ergoabcde}. 
Points {\bf a} and {\bf e} are in the region where HBHs have ergo-spheres. Point {\bf c} falls into the region where ergo-Saturns exist (shaded in light blue). This region is delimited by two lines that meet at a cusp (point {\bf 2}): the first line, connecting points {\bf 1} and {\bf 2}, corresponds to the appearance/disappearance of an ergo-torus -- which has zero radius along this line -- disconnected from the ergo-sphere. The second one, where point {\bf d} stands, corresponds to the merger/separation between an ergo-torus with non-zero radius and the ergo-sphere. Precisely at a cusp, the merging/separating ergo-torus has zero radius. The inset shows that the second line asymptotes to the boson star line.} 
\label{Mw}
\end{figure}

In Fig. \ref{Mw} we show the phase space of HBHs with $m=1$, even parity nodeless scalar fields~\cite{Herdeiro:2014goa} in a mass $M$ versus frequency $w$ diagram. As usual, we take $G=1$ and rescale quantities to the scalar field mass $\mu$. 

The phase space seen in Fig. \ref{Mw}   is delimited by three curves: the boson star curve (red solid line), the Kerr BHs curve (blue dotted line) and the extremal (i.e. zero temperature) HBHs curve (black dashed line). Along the boson star line, solutions starting with $M=0$ until point {\bf 1} are type 0; beyond point {\bf 1}  are type II. Moving from the boson star line to the interior of the HBHs domain of existence (shaded region), we find a line of solutions starting from point {\bf 1}  going through a cusp at  point {\bf 2} and approaching asymptotically the boson star line as one moves to the center of the spiral. This line separates type I and type III HBHs. Type I HBHs connect to type 0 boson stars whereas type III HBHs connect to type I boson stars. The transition from type I and type III HBHs can occur in two qualitatively distinct ways. Along the line segment connecting points {\bf 1} and {\bf 2} an ergo-torus appears/disappears, disconnected from the ergo-sphere. By contrast, along the line starting at point {\bf 2}, an ergo-torus pinches off from the ergo-sphere. The cusp, i.e point {\bf 2} corresponds to the limit when the ergo-torus pinching off has zero radius. 

In Fig. \ref{ergoabcde}  we show a cross section of the ergo-surface(s) for the 5 points highlighted in Fig. \ref{Mw}, that lie on a constant $q$ curve. Solutions {\bf a}, {\bf c} and {\bf e} are, respectively type I, III and I. Solutions {\bf b} and {\bf d} show the two possible transitions between type I and type III, as described above.

\newpage

\begin{widetext}

\begin{figure}[h!]
\centering
\includegraphics[height=1.69in]{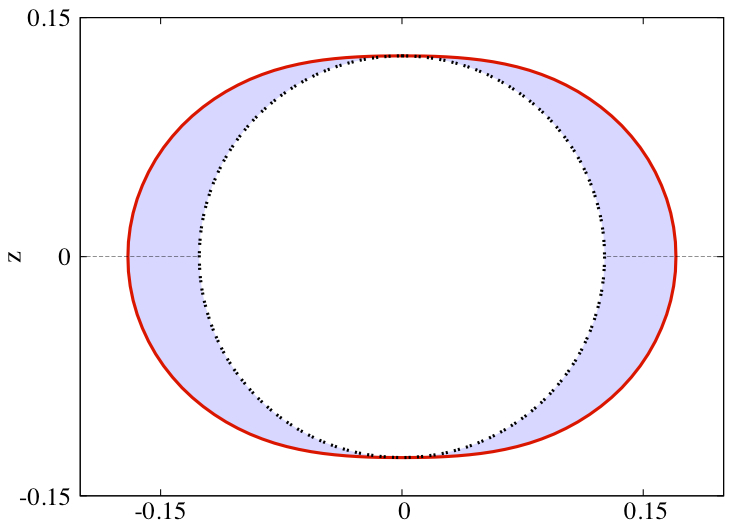}\ \ \ \ \ 
\includegraphics[height=1.69in]{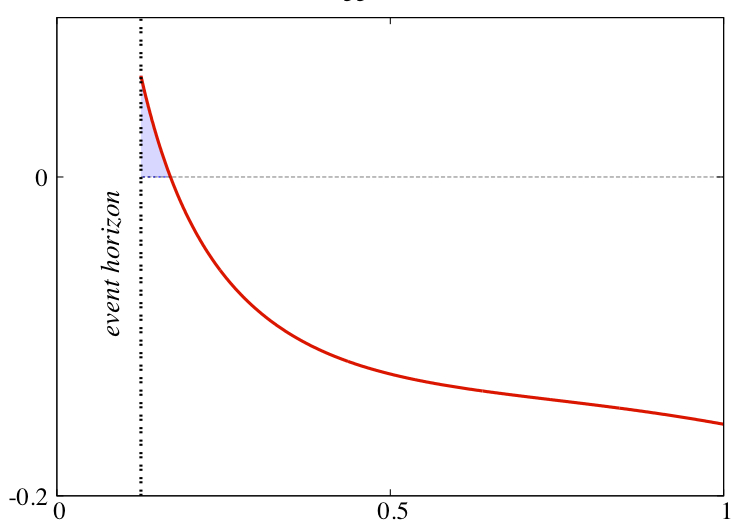}\\
\includegraphics[height=1.69in]{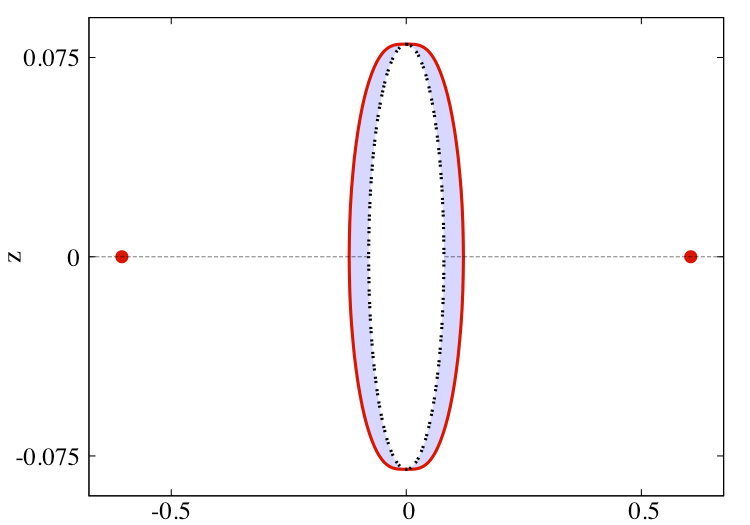}\ \ \ \ \ 
\includegraphics[height=1.69in]{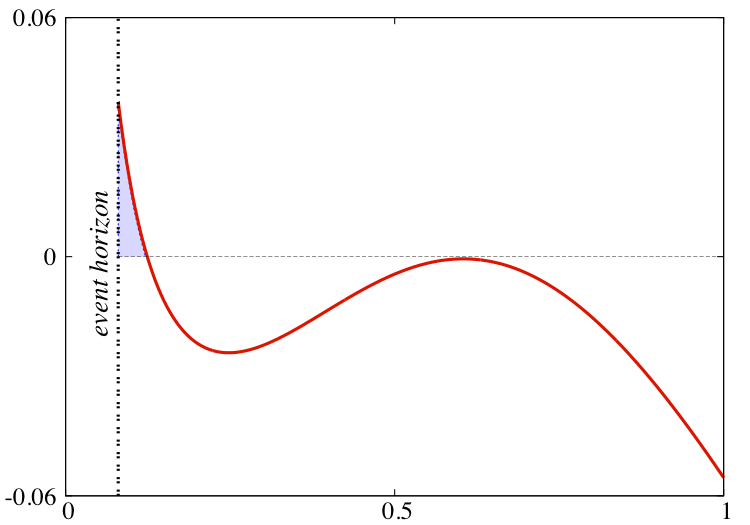}\\
\includegraphics[height=1.69in]{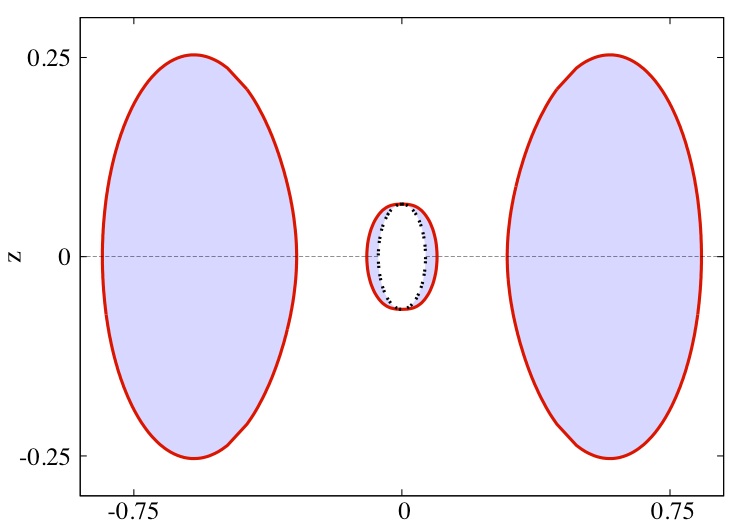}\ \ \ \ \ 
\includegraphics[height=1.69in]{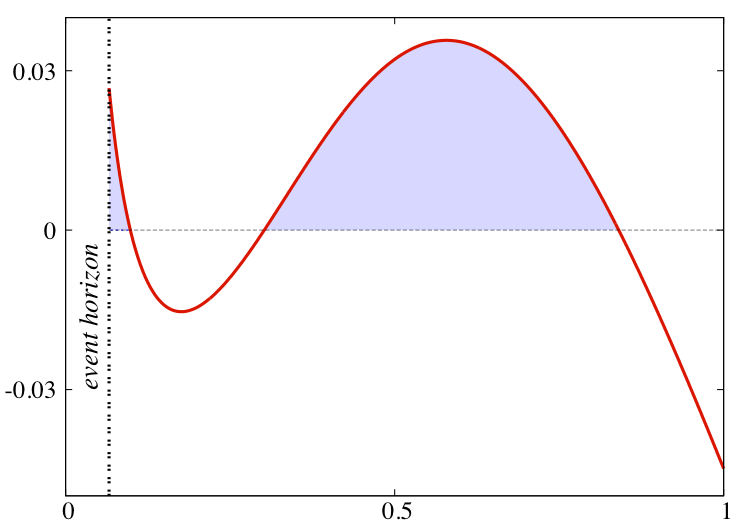}\\
\includegraphics[height=1.69in]{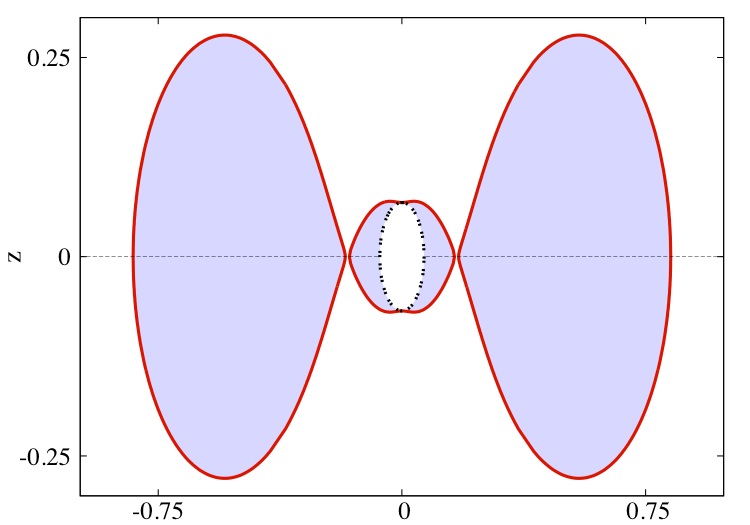}\  \ \ \ \ 
\includegraphics[height=1.69in]{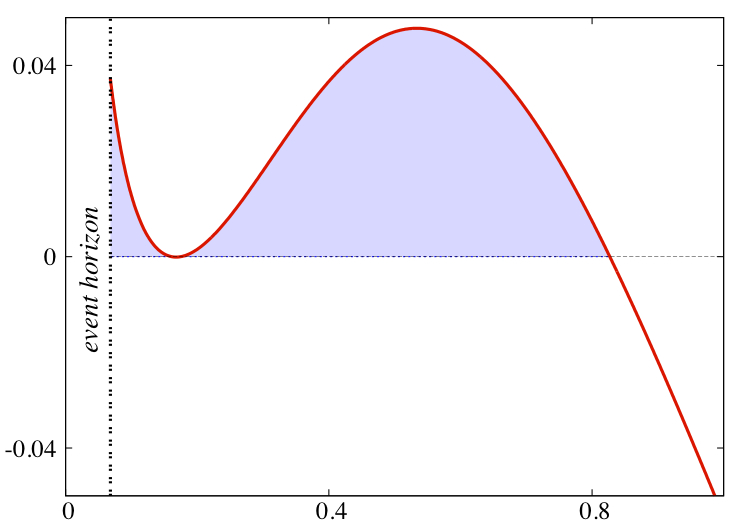}\\
\includegraphics[height=1.69in]{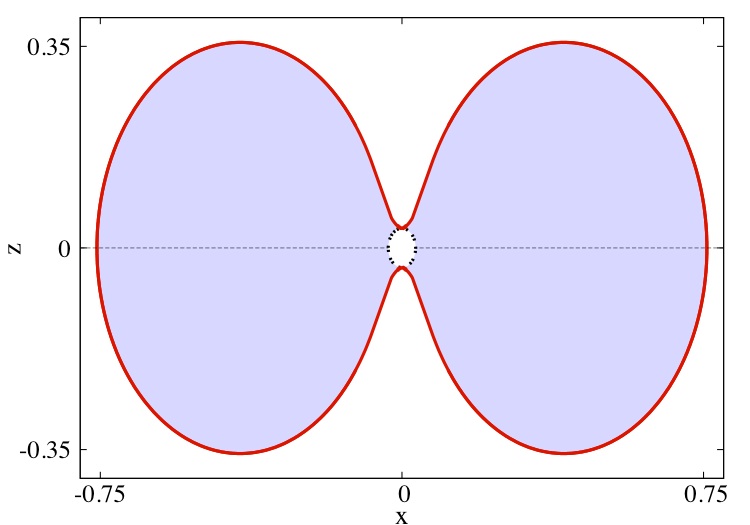}\ \ \ \ \ 
\includegraphics[height=1.69in]{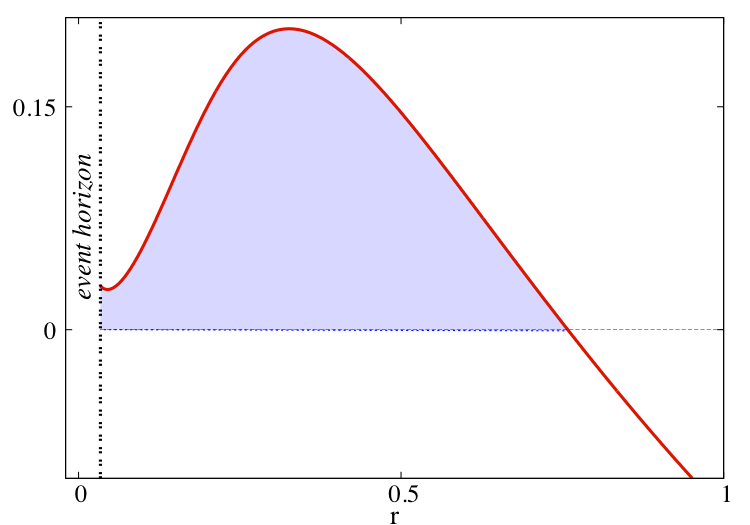}\\
\caption{A cross section of the ergo-surface(s) along the $x-z$ plane (left panel) and the metric function $g_{tt}$ along the equatorial plane (right panel) for the HBHs at points {\bf a} (top panel) to {\bf e} (bottom panel) in Fig. \ref{Mw}. 
%The event horizon is displayed as the black dotted curve and the ergo surface(s) as the solid red line(s). The ergo-region is shaded in blue. The top panel shows a type I HBH with an ergo-sphere. The bottom panel shows a type III HBH, with an ergo-Saturn. The transition between the two solutions occurs via the appearance of a zero radius ergo-torus disconnected from the ergo-sphere -- shown in the middle panel.
} 
\label{ergoabcde}
\end{figure}

%\begin{figure}[h!]
%\centering
%%\includegraphics[height=2.45in]{ergo-d.eps}\ 
%%\includegraphics[height=2.45in]{gtt-d.eps}\\
%\includegraphics[height=2.45in]{ergo-e.eps}\
%\includegraphics[height=2.45in]{gtt-e.eps}\\
%\caption{Same representations as in Fig. \ref{ergoabc} for the HBHs at points {\bf d} (top panel) and {\bf e} (bottom panel) in Fig. \ref{Mw}. The bottom panel again represents type I HBHs. But the transition to the type III HBHs represented in the bottom panel of Fig. \ref{ergoabc} is now done through the pinching off of an ergo-torus -- top panel.} 
%\label{ergode}
%\end{figure}
%

\end{widetext}

In Fig. \ref{ergoabcde}, left panels, the event horizon is displayed as the black dotted curve and the ergo-surface(s) as the solid red line(s). The ergo-region is shaded in blue. The top and bottom panels show type I HBHs with an ergo-sphere (topologically). The middle panel shows a type III HBH, with an ergo-Saturn. The first possible transition between the two types of HBH solutions occurs via the appearance of a zero radius ergo-torus disconnected from the ergo-sphere -- shown in the second panel. The second possible transition is through the pinching off of an ergo-torus, shown in the fourth panel. As for the 3D plots, the coordinates in equation (\ref{ansatz}) are plotted as standard spherical coordinates and, in particular, define the $x-z$ plane as standard Cartesian coordinates.

%\newpage

%%%%%%%%%%%%%%%%%
\section{Ergo-size and superradiant instability time-scale}
\label{stability}
\begin{figure}[h!]
\centering
\includegraphics[height=2.48in]{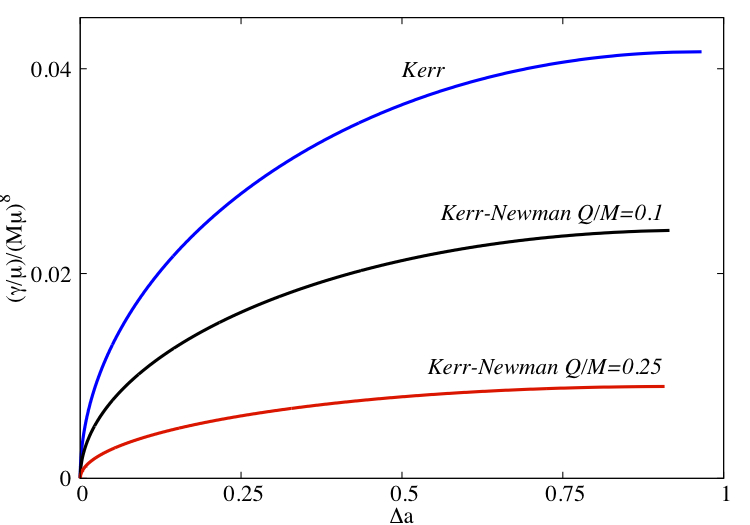}\\
\caption{Strength of the superradiant instability as a function of the ergo-size $\Delta a$, fixing $M$, $\mu$ and, in the charged case, $Q$. The instability grows monotonically with the ergo-size for all cases. These results are valid for $\mu M\ll 1$ and, in the charged case, $|qQ|\ll 1$.} 
\label{tau}
\end{figure}

%%%%%%%%%%%%%%%%%
Consider a complex, massive Klein-Gordon field, with mass $\mu$, in the background of a Kerr BH with mass $M$ and angular momentum $J$. For a mode with frequency $\omega$ in the superradiant regime, the strength of the instability is measured by the imaginary part of $\omega$, here denoted as $\gamma$. The inverse of $\gamma$ is the e-folding time scale for the instability, $\tau$.  It was shown by Detweiler~\cite{Detweiler:1980uk} that  $\tau$ in the limit when $M\mu\ll 1$, is given by:
\bequ
\tau \mu=\frac{24}{(\mu M)^8}\frac{M^2}{J} \ .
\label{tau_small}
\eequ
In this limit, fixing $M$ and $\mu$, we therefore see that the instability becomes monotonically stronger -- i.e. the time-scale decreases -- with increasing $J$. Alternatively, we can regard how $\gamma=1/\tau$ varies with the \textit{ergo-size}, as measured by $\Delta a$ defined in equation (\ref{ergosize}). Not surprisingly, we find that the strength of the instability increases monotonically with the ergo-size - Fig. \ref{tau}. We remark that the same qualitative behaviour is found in the Kerr-Newman family. Taking a scalar field minimally coupled to the electromagnetic field, with gauge coupling $q$, in the limit of $\mu M\ll1$ and $|qQ|\ll 1$, where $Q$ is the BH charge, it is found~\cite{Furuhashi:2004jk}:
\bequ
\tau \mu=\frac{24}{\mu^3|\mu M -qQ|^5}\frac{M}{J(M^2-Q^2)} \ .
\eequ
Fixing $q=\mu$, as a representative example for our purpose, and taking two different non-zero values of $Q$, we have also plotted the dependence of the strength of the instability in terms of the ergo-size in Fig. \ref{tau}, observing the same qualitative behaviour as for Kerr.\footnote{For a charged scalar field in the background of an asymptotically flat, non-rotating, charged BH there can be superradiant \textit{scattering}, extracting Coulomb energy and charge from the BH. There is, however, no superradiant \textit{instability}~\cite{Hod:2013nn,Hod:2013eea}, unless a mirror boundary condition for the scalar field is imposed~\cite{Herdeiro:2013pia,Hod:2013fvl,Degollado:2013bha}, or in $AdS$ asymptotics (see e.g.~\cite{Wang:2014eha} and references therein). As such, we may expect a correlation between the ergo-size and the strength of the instability for rotating BHs  even if charge is included, as in~\cite{Furuhashi:2004jk}.}

We would like to emphasize that the relation between the ergo-size and the strength of superradiant instabilities is heuristic. Moreover, the ergo-size is blind to the scalar field parameters. As such, even if it may be informative in the limit in which the Compton wave length of the particle is large enough to probe the ergo-size, it will certainly fail in the limit of very small Compton wave length. 

Actually, the ergo-size seems to stop being valid as a measure of the strength of the instability in the regime $\mu M\sim 1$. This can be justified with the results in~\cite{Dolan:2012yt}. There, it was observed that, in  $M\mu\sim 1$ regime, the strongest instability for fixed $M\mu$ does not occur precisely at extremality, but slightly before: a maximal growth rate of $M\gamma\sim 1.72 \times 10^{-7}$ was obtained for $\mu M\sim 0.45$ and $J/M^2\sim 0.997$. We shall therefore take that the ergo-size is a valid measure of the instability only in the regime $\mu M\ll 1$.

\begin{figure}[h!]
\centering
\includegraphics[height=2.48in]{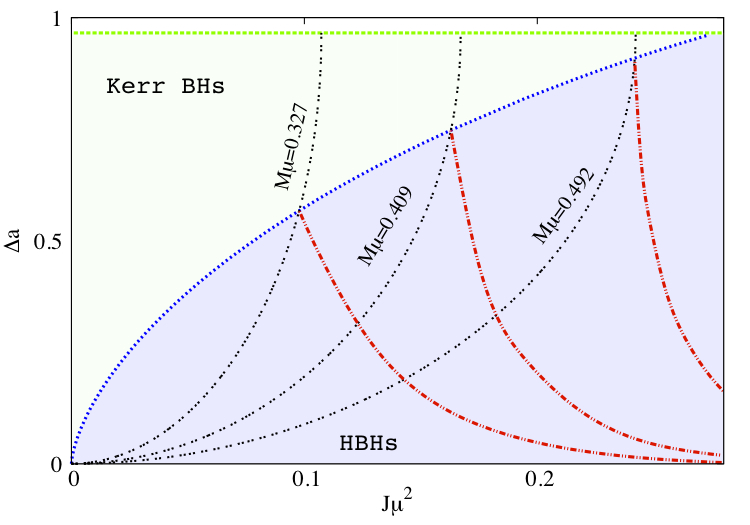}\\
\caption{Ergo-size $\Delta a$ as a function of the ADM angular momentum (in terms of the field mass) in the region of non-uniqueness. Kerr BHs exist below the solid green line which corresponds to extreme Kerr. HBHs exist in the shaded area below the dotted blue line, which corresponds to the Kerr limit of HBHs. Iso-mass curves are plotted for 3 different values of $M\mu$ both for Kerr BHs (dotted black lines) and HBHs (dashed red lines).} 
\label{deltaa}
\end{figure}

In Fig. \ref{deltaa} we show the ergo-size versus the ADM angular momentum for HBHs, in the region of non-uniqueness. The figure shows that for the same ADM mass and angular momentum, the ergo-size is smaller for HBHs. This is intuitive if one regards HBHs as bound states of boson stars -- which have no ergo-surfaces in this region of parameter space -- and Kerr BHs. Observe that the pattern observed for the iso-mass curves plotted also holds for $\mu M\ll 1$.

In the regime where eq. (\ref{tau_small}) is valid, the time-scale for the instability of Kerr BHs has the lower bound, in physical units, $\tau\sim 0.12(\mu M)^{-9}(M/M_\odot)\,  {\rm ms}$. According to the argument given above we expect this to be a lower limit for the e-folding time scale of superradiant instabilities for HBHs, in the region of non-uniqueness, where $\mu M\ll 1$. For instance, if $\mu M\sim 0.1$ this time scale varies from $\tau \sim 1\ {\rm day}$ to $\tau \sim 10^6 \ {\rm years}$, for solar mass BHs and the largest supermassive BHs known (with $M\sim 10^9 M_{\odot}$), respectively. 

Another perspective on the estimate we have just done is the following. Bench marking the time-scale for superradiant instabilities of HBHs using Kerr solutions is certainly well justified in the region where HBHs are close to Kerr, i.e. on a band next to the blue dotted line in Fig. \ref{Mw}. On such band, the scalar field should be `small' and hence the test field approximation should be legitimate. The `smallness' of the scalar field in this region also implies that the estimate should be universal: it should apply equally to scalar field models including self-interactions,  as non-linearities should be negligible. Finally, observe that for $\mu M \ll 1$ this band appears to encompass most, if not all, HBHs.

%%%%%%%%%%%%%%%%%
\section{Further Remarks}
\label{discussion}
%%%%%%%%%%%%%%%%%
In this paper we have analyzed the ergo-regions of the HBHs found in~\cite{Herdeiro:2014goa}. These solutions display an interesting and novel structure of ergo-surfaces and, in particular, provide the first four dimensional example where ergo-Saturns exist. We expect that some features concerning ergo-surfaces described herein will be present in other models of BHs with `hair'. For example, ergo-Saturns should also exist in the case when the scalar field
possesses a  potential with self-interaction terms.
%In this case,
%an interesting feature in this case is that the solitons 
%do not trivialize in the flat spacetime limit, becoming the Q-balls introduced by Coleman in [1].
%Indded, the results in \cite{Kleihaus:2007vk}
%show that a spinning gravitating Q-ball 
%possess also a toroidal ergoregion, in some region of the parameter space.
This is the case for the
HBHs generalization of the spinning gravitating Q-balls
in \cite{Kleihaus:2005me}. Indeed, for this model, we have found solutions with rather similar properties
to those discussed above \cite{Herdeiro:2014}.

Another, more involved, version
of the same picture is provided by BHs inside vortons.
Vortons~\cite{Radu:2008pp,Garaud:2013iba}
may be regarded as four dimensional field theory analogues of the higher dimensional black rings of vacuum general relativity \cite{Emparan:2001wn}.  They are
made from loops of vortices, which are balanced against collapse by rotation,
and share many features of the Q-balls.
In particular, as shown in \cite{Kunz:2013wka}, gravitating vortons may posses an ergo-region with a toroidal shape. Thus their BH generalization will also possess 
 ergo-Saturns  \cite{Herdeiro:2014}.
 
 In fact, it is reasonable to expect that the occurence of ergo-Saturns will be a feature of any theory which
 possess spinning soliton solutions in the flat spacetime limit.
%The arguments in Section II are rather general: 
When including  gravity effects, these solitons
should  develop an ergo-torus and, when a rotating BH can be added at the centre of the spinning soliton, an ergo-sphere will also be present. The upshot will be the presence of an ergo-Saturn. 
%However, it should always be possible 
%to add a small black hole at the center of any
%spinning soliton.
%Then the resulting HBH would possess an ergo-sphere
%or an ergo-Saturn.
%\footnote{Indeed,
%this is the case for all known solutions,
%see e.g.$

Even more diverse ergo-surfaces should be found by considering $composite$, rather than single, solitons.
As an example take the same ansatz used herein -- equations (\ref{ansatz}) --,
but considering instead an \textit{odd-parity} scalar field\footnote{Observe, however, that even for these odd-parity solutions the metric functions will be invariant under a reflection along the equatorial plane.}.
Axially symmetric spinning boson stars can be found  
describing composite configurations
with double torus -- i.e. 
dumbbell-like -- surfaces of constant energy density  \cite{Kleihaus:2007vk}.
%In this case,
%the energy-momentum and charge densities show two maxima located symmetrically
%with respect to the equatorial plane.  
%the configuration being prevented from collapse by the spin-spin interaction.
%The results in \cite{Kunz:2013wka}
% show that, 
As such, in some region of the parameter space, these negative parity solitons
develop an ergo-region
consisting of two disconnected tori, 
located symmetrically
with respect to the equatorial plane. 
We anticipate the existence of HBH generalizations of these boson stars,
with an ergo-region consisting on three disconnected parts:
two ergo-tori located symmetrically on the symmetry axis
and a central ergo-sphere.
%
%\footnote{Note that we  consider
%a single black hole only,
% located
%at the symmetry center of the configuration.
%However, one cannot $apriori$ exclude
%the possibility that the extra-interactions introduced by the matter fields
%would balance a multi-Kerr black hole solution.
%} 
%apart from BHs with a spherical ergo-region. 
%Close to the merger between the ergo-tori and the ergo-sphere it is even possible that voids exist, i.e.  ergo-regions inside which 
%cannot be excluded, especially close to the transition regions.
%

Even more involved ergo-regions are expected to exist for 
spinning HBHs in models possessing  $chain$ solitons, with more than two 
individual components located on the symmetry axis 
(see
$e.g.$ 
\cite{Kleihaus:2003nj,Ibadov:2010hm}). In some regions of the parameter space, the
ergo-surfaces of the HBHs obtained from these solutions
would possess more than three distinct components,
with an intricate transition scenario.  

Yet a qualitatively different case is found in higher dimensions. In five spacetime dimensions, asymptotically flat boson stars have been found with an \textit{ergo-shell}~\cite{Hartmann:2010pm}; this is a topologically spherical ($S^3$) hollow ergo-region. We have constructed solutions in which a BH is superimposed at the centre of such boson stars. Some of these HBH solutions, which will be reported elsewhere, have an ergo-sphere inside the hollow ergo-shell.

Let us conclude with a remark concerning the superradiant instabilities that are likely to be present for the HBHs analyzed in this paper. 
Even if such instabilities are confirmed, it should be clear that this fact, \textit{per se}, does not rule out that HBHs may play a physical role. Unstable configurations play important roles in nature; an obvious example are unstable nuclei, such as Uranium or Plutonium. The crucial point is the time-scales of the instabilities.

The estimate we have put forward for the e-folding time, applying to a particular region of the HBHs phase space, suggests that the decay time may be large enough for the `hairiness' to become of relevance in physical processes. In this respect, observe that the time-scales estimated from the Kerr solutions  were computed from the fastest growing mode, i.e the mode with angular momentum harmonic index $\ell=1$. For the HBHs discussed here, with $m=1$, it is expected that only modes with $m>1$ can give rise to an instability, which implies $\ell>1$. Since it is known that the strength of the instability decays with increasing $\ell$~\cite{Detweiler:1980uk,Furuhashi:2004jk}, the time-scales we discussed provide a conservative lower bound for the e-folding time of the possible superradiant instabilities.

\bigskip

%Beautiful analysis of ergoregions merger~\cite{Elvang:2008qi}

%REF FOR Invariant definition ergospheres

%%%%%%%%%%%%%%%%%%%%%%%%%%%%%%%%%%%%%%%%%%%%%%%
\section*{Acknowledgements}
%%%%%%%%%%%%%%%%%%%%%%%%%%%%%%%%%%%%%%%%%%%%%%%
The authors are supported by the FCT IF program. The work in this paper is also supported by the grants PTDC/FIS/116625/2010 and  NRHEP--295189-FP7-PEOPLE-2011-IRSES.

\bibliographystyle{h-physrev4}
\bibliography{ergoregions}

%\newpage

\end{document}